\documentstyle [12pt] {article}

\catcode`@=12
\begin{document}
\thispagestyle{empty}
\begin{flushright}
UCRHEP-T103\\
TRI-PP-93-3\\
January 1993\
\end{flushright}
\vspace{0.5in}
\begin{center}
{\Large \bf Symmetry Remnants: Rationale for Having\\
Two Higgs Doublets\\}
\vspace{1.0in}
{\bf Ernest Ma\\}
\vspace{0.1in}
{\sl Department of Physics\\}
{\sl University of California\\}
{\sl Riverside, California 92521\\}
\vspace{0.3in}
{\bf Daniel Ng\\}
\vspace{0.1in}
{\sl TRIUMF, 4004 Wesbrook Mall\\}
{\sl Vancouver, British Columbia\\}
{\sl Canada V6T 2A3\\}
\vspace{1.0in}
\end{center}
\begin{abstract}\
There is a good reason why the standard electroweak
${\rm SU(2) \times U(1)}$ gauge model may be supplemented by two
Higgs scalar doublets.  They may be remnants of the spontaneous breaking
of an ${\rm SU(2) \times SU(2) \times U(1)}$ gauge symmetry at a much
higher energy scale.  In one case, the two-doublet Higgs potential has a
custodial SU(2) symmetry and implies an observable scalar triplet.  In
another, a light neutral scalar becomes possible.
\end{abstract}
\newpage
\baselineskip 24pt

In the standard ${\rm SU(2) \times U(1)}$ electroweak gauge model,
only one Higgs scalar
doublet is needed for the spontaneous generation of all particle masses.  Yet
there are numerous research papers dealing with the possibility of having two
(or more) doublets.\cite{ghkd}  A good reason is of course supersymmetry, but
in that case, there should be many other particles as well.  Nevertheless, a
general two-doublet extension of the standard electroweak model without
supersymmetry is routinely studied with little theoretical justification other
than the obvious fact that it is not known to be wrong.  To remedy this
situation, we will show in the following that if the standard
${\rm SU(2) \times U(1)}$
electroweak gauge group is the remnant of a larger symmetry, then the
appearance of two (or more) doublets at the electroweak energy scale is
actually required in some cases and the special form of the corresponding
Higgs potential may even be indicative of what the larger theory is.

Consider the following Higgs potential for two doublets:\cite{fmk2}
\begin{eqnarray}
V &=& \mu_1^2 \Phi_1^\dagger \Phi_1 + \mu_2^2 \Phi_2^\dagger \Phi_2 +
\mu_{12}^2 (\Phi_1^\dagger \Phi_2 + \Phi_2^\dagger \Phi_1) \nonumber \\
&+& {1 \over 2} \lambda_1 (\Phi_1^\dagger \Phi_1)^2 + {1 \over 2} \lambda_2
(\Phi_2^\dagger \Phi_2)^2 + \lambda_3 (\Phi_1^\dagger \Phi_1)(\Phi_2^\dagger
\Phi_2) \nonumber \\ &+& \lambda_4 (\Phi_1^\dagger \Phi_2)(\Phi_2^\dagger
\Phi_1) + {1 \over 2} \lambda_5 (\Phi_1^\dagger \Phi_2)^2 + {1 \over 2}
\lambda_5^* (\Phi_2^\dagger \Phi_1)^2,
\end{eqnarray}
where
\begin{equation}
\Phi_{1,2} = \left( \begin{array} {c} \phi_{1,2}^+ \\ \phi_{1,2}^0 \end{array}
\right)
\end{equation}
and $\mu_{12}^2$ has been chosen real by virtue of the arbitrary phase between
$\Phi_1$ and $\Phi_2$.  This $V$ is invariant under a $Z_2$ discrete symmetry
where $\Phi_1~(\Phi_2)$ may be considered even (odd) except for the
$\mu_{12}^2$ term, but which breaks it only softly.  Consequently, it allows
for the natural suppression of flavor-changing neutral currents as long as
each fermion gets its mass from only one scalar vacuum expectation value,
{\it i.e.} either $<\phi_1^0>$ or $<\phi_2^0>$ but not both.

Such two-doublet extensions of the standard electroweak model have been
studied extensively for their phenomenological implications.  However,
a more fundamental question to be considered is why they should be studied
at all.  In supersymmetry, two scalar doublets are necessary because each
is accompanied by a fermionic partner having a nonzero contribution to
the axial-vector triangle anomaly but their sum is zero.  The requirement of
supersymmetry also constrains the parameters of $V$ as follows:
\begin{equation}
\lambda_1 = \lambda_2 = {1 \over 4} (g_1^2 + g_2^2),~~\lambda_3 = -{1 \over 4}
g_1^2 + {1 \over 4} g_2^2,~~\lambda_4 = -{1 \over 2} g_2^2,~~
\lambda_5 = 0,
\end{equation}
where $g_1$ and $g_2$ are the U(1) and SU(2) gauge couplings of the
standard model respectively.  The soft terms, {\it i.e.} $\mu_1^2$, $\mu_2^2$,
and $\mu_{12}^2$, are considered arbitrary because they are allowed to break
the supersymmetry.  Discovery of scalar particles with a mass spectrum
conforming to such a Higgs potential would certainly be a strong indication
of supersymmetry.

Consider now a different rationale for the existence of two Higgs doublets.
They may be remnants of the spontaneous breaking of a larger gauge symmetry
at some higher energy scale.  Take for example the gauge group
${\rm SU(2)_1 \times SU(2)_2 \times U(1)}$.
Let the scalar sector consist of two doublets $\Phi_{1,2}$
and one self-dual bidoublet $\eta$ transforming as (2,1,1/2), (1,2,1/2), and
(2,2,0) respectively:
\begin{equation}
\Phi_{1,2} = \left( \begin{array} {c} \phi_{1,2}^+ \\ \phi_{1,2}^0
\end{array} \right),~~\eta = {1 \over \sqrt 2} \left( \begin{array}
{c@{\quad}c} \overline {\eta^0} & \eta^+ \\ -\eta^- & \eta^0 \end{array}
\right).
\end{equation}
The most general Higgs potential $V$ is then given by
\begin{eqnarray}
V &=& m_1^2 \Phi_1^\dagger \Phi_1 + m_2^2 \Phi_2^\dagger \Phi_2 +
m_3^2 Tr (\eta^\dagger \eta) \nonumber \\ &+& {1 \over 2} f_1 (\Phi_1^\dagger
\Phi_1)^2 + {1 \over 2} f_2 (\Phi_2^\dagger \Phi_2)^2 + {1 \over 2} f_3
(Tr (\eta^\dagger \eta))^2 \nonumber \\ &+& f_4 (\Phi_1^\dagger \Phi_1)
Tr (\eta^\dagger \eta) + f_5 (\Phi_2^\dagger \Phi_2) Tr (\eta^\dagger \eta) +
f_6 (\Phi_1^\dagger \Phi_1) (\Phi_2^\dagger \Phi_2) \nonumber \\ &+&
t (\Phi_1^\dagger \eta \Phi_2 + \Phi_2^\dagger \eta^\dagger \Phi_1),
\end{eqnarray}
where $t$ has been chosen real by virtue of the arbitrary relative phase
between $\Phi_1$ and $\Phi_2$.  Note that because
\begin{equation}
\Phi_1^\dagger \eta \Phi_2 + \Phi_2^\dagger \eta^\dagger \Phi_1 = Tr \left(
\begin{array} {c@{\quad}c} \phi_1^0 & -\phi_1^+ \\ \phi_1^- & \overline
{\phi_1^0} \end{array} \right) {1 \over \sqrt 2} \left( \begin{array}
{c@{\quad}c} \overline {\eta^0} & \eta^+ \\ -\eta^- & \eta^0 \end{array}
\right) \left( \begin{array} {c@{\quad}c} \overline {\phi_2^0} & \phi_2^+
\\ -\phi_2^- & \phi_2^0 \end{array} \right),
\end{equation}
this $V$ has automatically an extra global SU(2) symmetry.\cite{bwm}  As the
first step of symmetry breaking, consider only $<\eta^0> = <\overline {\eta^0}>
= u \neq 0$, then our ${\rm SU(2)_1 \times SU(2)_2 \times U(1)}$ breaks down
to the standard ${\rm SU(2)_L \times U(1)_Y}$,
resulting in a massive vector-boson
triplet $(g_1 W_1^{\pm,0} - g_2 W_2^{\pm,0})/\sqrt {g_1^2+g_2^2}$ and
preserving the extra global SU(2) symmetry.  The reduced $V$ now has the
form of Eq. (1) but with the important restriction that $\lambda_4 =
\lambda_5 = 0$.  [The other parameters are $\mu_1^2 = m_1^2 + f_4 u^2$,
$\mu_2^2 = m_2^2 + f_5 u^2$, $\mu_{12}^2 = tu/\sqrt 2$, $\lambda_1 = f_1$,
$\lambda_2 = f_2$, and $\lambda_3 = f_6$.]  Both $\Phi_1$ and $\Phi_2$
now transform as doublets under the standard
${\rm SU(2)_L \times U(1)_Y}$ gauge
group, as well as the extra global SU(2).  As $\phi_1^0$ and $\phi_2^0$
acquire vacuum expectation values $v_1$ and $v_2$, the gauge symmetry
${\rm SU(2)_L \times U(1)_Y}$
breaks down to electromagnetic ${\rm U(1)_Q}$, but a
custodial SU(2) symmetry remains, in exact analogy to the well-known
case of the standard model with only one Higgs doublet.  Consequently,
of the 5 physical scalar bosons, 3 are organized into a triplet
\begin{eqnarray}
H_3^\pm &=& -\sin \beta \phi_1^\pm + \cos \beta \phi_2^\pm, \\
H_3^0 &=& \sqrt 2 (-\sin \beta Im \phi_1^0 + \cos \beta Im \phi_2^0),
\end{eqnarray}
where $\tan \beta \equiv v_2/v_1$, with a common mass given by
\begin{equation}
m_{H_3}^2 = {{-2 \mu_{12}^2} \over {\sin 2 \beta}}.
\end{equation}
The other 2 are singlets
\begin{eqnarray}
H_1 &=& \sqrt 2 (\cos \beta Re \phi_1^0 + \sin \beta Re \phi_2^0), \\
H_2 &=& \sqrt 2 (-\sin \beta Re \phi_1^0 + \cos \beta Re \phi_2^0),
\end{eqnarray}
with mass-squared matrix given by
\vspace {0.1in}
\begin{equation}
{\cal M}^2 = \left( \begin{array} {c@{\quad}c} 2(c^2 \lambda_1 v_1^2 +
s^2 \lambda_2 v_2^2 + 2 s c \lambda_3 v_1 v_2) & 2 s c ((-\lambda_1 +
\lambda_3) v_1^2 + (\lambda_2 - \lambda_3) v_2^2) \\ 2 s c ((-\lambda_1
+ \lambda_3) v_1^2 + (\lambda_2 - \lambda_3) v_2^2) & m_{H_3}^2 + 2 s c
(\lambda_1 + \lambda_2 -2 \lambda_3) v_1 v_2 \end{array} \right),
\end{equation}
\vspace{0.1in}
where $s = \sin \beta$, $c = \cos \beta$.  Since $\lambda_1 + \lambda_2 >
2|\lambda_3|$ is required for $V$ to be bounded from below, the above
matrix shows that at least one of the singlet scalars must be heavier
than the triplet.

The $V$ of Eq. (1) is in general not invariant under an extra global
SU(2) symmetry, hence the presence of two Higgs doublets is expected
to contribute significantly to the radiative correction which makes the
electroweak parameter $\rho$ different from one.\cite{fmk1}  Experimentally,
there is no evidence of any deviation which cannot be accounted for with a
$t$-quark mass of about 150 GeV or so.  Hence such a custodial symmetry is
desirable for $V$, but that would require\cite{dm} $\lambda_4 = \lambda_5$
which cannot be maintained naturally in the context of the standard model
because infinite radiative corrections are unavoidable.  In our case, the
restriction $\lambda_4 = \lambda_5 = 0$ is obtained from the reduction
of a larger theory and it can easily be shown that $\lambda_4$ and
$\lambda_5$ have finite radiative corrections which go to zero as $u$
goes to infinity.

The reason for both $\Phi_1$ and $\Phi_2$ to be present in the reduced Higgs
potential has to do with the original
${\rm SU(2)_1 \times SU(2)_2 \times U(1)}$ theory.
If some of the fermions couple to ${\rm SU(2)_1 \times U(1)}$
and others to ${\rm SU(2)_2 \times U(1)}$,
then both $\Phi_1$ and $\Phi_2$ are required to allow all
fermions to acquire mass.\cite{lm}  At the $10^2$ GeV energy scale, all
fermions couple to the standard ${\rm SU(2) \times U(1)}$
in the usual way and the only
clue to their original difference is the two Higgs doublets with $\lambda_4 =
\lambda_5 = 0$ in $V$.  Discovery of the scalar triplet $H_3^{\pm,0}$
would certainly be indicative of such a possibility.

As a second example, consider again the gauge group
${\rm SU(2)_1 \times SU(2)_2 \times U(1)}$
but with an unconventional assignment of fermions.\cite{metc}  An exotic quark
$h$ of electric charge $-1/3$ is added so that $(u,d)_L$ transforms as
(2,1,1/6), $(u,h)_R$ as (1,2,1/6), whereas both $d_R$ and $h_L$ are singlets
(1,1,$-1/3$).  There are again the two Higgs doublets $\Phi_{1,2}$ but now
the bidoublet is not self-dual, {\it i.e.}
\begin{equation}
\eta = \left( \begin{array} {c@{\quad}c} \overline {\eta_1^0} & \eta_2^+ \\
-\eta_1^- & \eta_2^0 \end{array} \right).
\end{equation}
\vspace{0.1in}
Assume also a $Z_4$ discrete symmetry under which $\Phi_1 \rightarrow \Phi_1$,
$\Phi_2 \rightarrow i\Phi_2$, $\eta \rightarrow i\eta$, $(u,d)_L \rightarrow
(u,d)_L$, $(u,h)_R \rightarrow -i(u,h)_R$, $d_R \rightarrow d_R$, and
$h_L \rightarrow -h_L$.  This then forces $h_L$ to pair up with $h_R$ via
$<\phi_2^0> = v_2$, $d_L$ with $d_R$ via $<\phi_1^0> = v_1$, and $u_L$ with
$u_R$ via $<\eta_1^0> = u_1$.  It also allows $<\eta_2^0> = 0$ as shown below.
The resulting theory retains an exact $Z_2$ discrete symmetry under which $h$,
$\eta_2$, and $W_R^\pm$ are odd and all the other particles are even.  It
can be thought of as a residual $R$-parity derived from the original
supersymmetric $E_6$ theory and has many interesting and remarkable
phenomenological consequences.\cite{others}

The most general Higgs potential $V$ invariant under the assumed $Z_4$
discrete symmetry is given by
\begin{eqnarray}
V &=& m_1^2 \Phi_1^\dagger \Phi_1 + m_2^2 \Phi_2^\dagger \Phi_2 + m_3^2
Tr(\eta^\dagger \eta) \nonumber \\ &+& {1 \over 2} f_1 (\Phi_1^\dagger
\Phi_1)^2 + {1 \over 2} f_2 (\Phi_2^\dagger \Phi_2)^2 + {1 \over 2} f_3
(Tr(\eta^\dagger \eta))^2 \nonumber \\ &+& {1 \over 4} f_4 Tr(\eta^\dagger
\tilde {\eta}) Tr(\tilde {\eta}^\dagger \eta) + {1 \over 8} f_5 (Tr(
\eta^\dagger \tilde {\eta}))^2 + {1 \over 8} f_5 (Tr(\tilde {\eta}^\dagger
\eta))^2 \nonumber \\ &+& f_6 (\Phi_1^\dagger \Phi_1) Tr(\eta^\dagger \eta)
+ f_7 (\Phi_2^\dagger \Phi_2) Tr(\eta^\dagger \eta) + f_8 (\Phi_1^\dagger
\eta \eta^\dagger \Phi_1) \nonumber \\ &+& f_9 (\Phi_2^\dagger
\eta^\dagger \eta \Phi_2) + f_{10} (\Phi_1^\dagger \Phi_1) (\Phi_2^\dagger
\Phi_2) + t (\Phi_1^\dagger \tilde {\eta} \Phi_2 + \Phi_2^\dagger \tilde
{\eta}^\dagger \Phi_1),
\end{eqnarray}
where
\begin{equation}
\tilde {\eta} \equiv \sigma_2 \eta^* \sigma_2 = \left( \begin{array}
{c@{\quad}c} \overline {\eta_2^0} & \eta_1^+ \\ -\eta_2^- & \eta_1^0
\end{array} \right).
\end{equation}
The couplings $f_5$ and $t$ have been chosen real by virtue of the
arbitrary relative phases among $\Phi_{1,2}$ and $\eta$.  As the first
step of symmetry breaking, consider now only $<\phi_2^0> = v_2 \neq 0$,
then ${\rm SU(2)_2 \times U(1)}$ breaks down to ${\rm U(1)_Y}$,
whereas ${\rm SU(2)_1}$ remains
unbroken and is in fact the standard ${\rm SU(2)_L}$.  Eliminating the heavy
$\Phi_2$ and $\eta_2$ scalar bosons from $V$, we again obtain Eq. (1)
but with $\lambda_5 = 0$ and $\Phi_2$ replaced by $\eta_1$.  [The other
parameters are $\mu_1^2 = m_1^2 + f_{10} v_2^2$, $\mu_2^2 = m_3^2 + f_7
v_2^2$, $\mu_{12}^2 = t v_2$, $\lambda_1 = f_1$, $\lambda_2 = f_3$,
$\lambda_3 = f_6 + f_8$, and $\lambda_4 = -f_8$.]  Note that the only
term in Eq. (14) involving 3 different neutral scalar fields is
$t (\overline {\phi_1^0} \eta_1^0 \phi_2^0 + \overline {\phi_2^0}
\overline {\eta_1^0} \phi_1^0)$ which means that $<\eta_2^0> = 0$ is
allowed.  Note also that because
$\eta$ is not self-dual, the $V$ of Eq. (14) does not have an extra
global SU(2) symmetry.  Hence $\lambda_4 \neq \lambda_5$ and $H_3^\pm$
and $H_3^0$ have different masses.  However, because $\lambda_5 = 0$, the
mass of $H_3^0$ is still given by Eq. (9), whereas
\begin{equation}
m_{H_3^\pm}^2 = m_{H_3^0}^2 - \lambda_4 (v_1^2 + u_1^2).
\end{equation}
Since $H_3^0$ is now the only scalar boson with a mass-squared proportional
to $\mu_{12}^2$, it may in fact be light.  [If $\mu_{12}^2$ were zero as
well as $\lambda_5$, then $V$ has an extra global U(1) symmetry, the
spontaneous breaking of which would result in a massless $H_3^0$.]  The
decay $Z^0 \rightarrow H_3^0 H_3^0$ is absolutely forbidden by
angular-momentum conservation and Bose statistics, whereas $Z^0 \rightarrow
H_{1,2}^0 H_3^0$ and $W^\pm \rightarrow H_3^\pm H_3^0$ may be forbidden
kinematically because $H_{1,2}^0$ and $H_3^\pm$ are heavy.  However, since
$H_{1,2}^0$ couple to $H_3^0 H_3^0$ through $V$, the decay $Z^0 \rightarrow
H_3^0 H_3^0 H_3^0$ may be possible, although the branching fraction is
expected to be very much suppressed.\cite{duma}  Note that $\lambda_5 = 0$
also in supersymmetry, but there it may be argued that $\mu_{12}^2$ should
not be small.  In the Yukawa sector, since $d_R$ only couples to $\Phi_1$
and $u_R$ only to $\eta_1$, the usual $Z_2$ discrete symmetry assumed for
the natural suppression of flavor-changing neutral currents is also
realized.

It has been shown in the above that the scalar sector accompanying the
standard model at the electroweak energy scale may very well consist of
two doublets, obeying the Higgs potential of Eq. (1), but with the
important restriction that $\lambda_4 = \lambda_5 = 0$ in the first case,
and $\lambda_5 = 0$ in the second.  These have interesting phenomenological
consequences because of the existence of an unbroken custodial SU(2)
symmetry in the former, and a softly broken U(1) symmetry in the latter.
A scalar triplet $H_3^{\pm,0}$ with a common mass is then predicted in
the first case, and a possibly light $H_3^0$ in the second.  Both can be
experimentally tested with future high-energy accelerators such as the
Superconducting Super Collider (SSC) and the Large Hadron Collider (LHC).

In closing, we should point out that with the fermionic content of our
second example, it is actually possible to have the same reduced $V$ as
in our first example, {\it i.e.} with $\lambda_4 = \lambda_5 = 0$, but
a rather {\it ad hoc} assumption is then required.  Let us choose the
bidoublet $\eta$ to be self-dual, which means that we cannot impose any
additional symmetry to distinguish $\eta$ from $\tilde \eta$ as in our
second example.  The mass matrix linking $(\overline {d}_L, \overline {h}_L)$
to $(d_R, h_R)$ is no longer restricted to be diagonal.  In particular,
there is a $\overline {h}_L d_R$ term.  However, if we make the {\it ad hoc}
assumption that this term is small compared to the $\overline {h}_L h_R$
term which comes from $<\phi_2^0>$, then again the heavy particles will
decouple and we obtain the $V$ of our first example.  Another way to
achieve this result is to forbid the $\overline {h}_L d_R$ term with a
discrete symmetry by adding a second $\Phi_2$, the existence of which is
of course not very well motivated.

The same $V$ for two Higgs doublets may come from very different models at
a much higher energy scale.  However, their couplings to the quarks and
leptons will generally not be the same.  We have not considered these
here because they are highly model-dependent.  If two Higgs doublets are
discovered in the future, detailed experimental determination of their
properties will likely point to a larger gauge theory at some higher
energy scale.
\vspace{0.3in}
\begin{center} {ACKNOWLEDGEMENT}
\end{center}

The work of E.M. was supported in part by the U. S. Department of Energy
under Contract No. DE-AT03-87ER40327.  The work of D.N. was supported by
the Natural Sciences and Engineering Research Council of Canada.

\newpage
\bibliographystyle{unsrt}

\end{document}